\documentclass[aps,twocolumn,nobibnotes,amsmath,amssymb]{revtex4-2}

\usepackage{xcolor}
\usepackage{hyperref}
\usepackage{orcidlink}
\definecolor{mycolor}{RGB}{0,88,204}
\hypersetup{
  colorlinks=true,
  linkcolor=black,
  urlcolor=mycolor,
  citecolor=mycolor
}
\usepackage{tikz}
\usetikzlibrary{calc,arrows.meta}
\definecolor{ladderInk}{HTML}{1A1A1A}
\definecolor{ladderSubtle}{HTML}{666666}
\definecolor{ladderVerif}{HTML}{1B7F5F}
\definecolor{ladderValid}{HTML}{C45A00}
\definecolor{ladderExp}{HTML}{5A4FB0}

\usepackage{mathptmx}
\DeclareMathAlphabet{\mathcal}{OMS}{cmsy}{m}{n}
\usepackage{listings}
\usepackage[T1]{fontenc}
\usepackage[utf8]{inputenc}
\usepackage{chngcntr}
\usepackage{inconsolata}
\usepackage{upquote}
\definecolor{keywordcolor}{rgb}{0.7, 0.1, 0.1}   
\definecolor{tacticcolor}{rgb}{0.0, 0.1, 0.6}    
\definecolor{commentcolor}{rgb}{0.4, 0.4, 0.4}   
\definecolor{symbolcolor}{rgb}{0.0, 0.1, 0.6}    
\definecolor{sortcolor}{rgb}{0.1, 0.5, 0.1}      
\definecolor{attributecolor}{rgb}{0.7, 0.1, 0.1} 

\lstnewenvironment{code}[1][]%
{
  \noindent\newline
  \minipage{1\linewidth}
  \vspace{0.5\baselineskip}
  \lstset{
    frame = lrtb,
    rulecolor=\color{mycolor},
    escapeinside={/*!}{!*/},
    language=lean,
    aboveskip = 5mm,
    belowskip = 5mm,
    xleftmargin=2mm,
    xrightmargin=2mm,
  }
}
{\endminipage\newline}

\usepackage{xpatch}
\usepackage{realboxes}
\definecolor{mycolorSubtle}{RGB}{245,250,255}
\DeclareRobustCommand{\myinline}{\lstinline}
\makeatletter
\xpretocmd\myinline{\Colorbox{mycolorSubtle}\bgroup\appto\lst@DeInit{\egroup}}{}{}
\makeatother
\begin{document}
\counterwithin{lstlisting}{section}

\title{The Future of AI-assisted Physics: Discovering, Verifying and Organizing Results via Formalization}

\author{Sven Krippendorf \orcidlink{0000-0001-6374-6828}}
\affiliation{Department of Applied Mathematics and Theoretical Physics, University of Cambridge, Wilberforce Road, Cambridge, CB3 0WA, U.K.}
\affiliation{Cavendish Laboratory, University of Cambridge, JJ Thomson Avenue, Cambridge, CB3 0HE, U.K.}
\email[Contact author: ]{slk38@cam.ac.uk}

\author{Leopoldo Sarra \orcidlink{0000-0001-7504-8656}}
\affiliation{Axiomatic AI, Barcelona, Spain}

\author{Joseph Tooby-Smith \orcidlink{0000-0003-2831-598X}}
\email[Contact author: ]{js4814@bath.ac.uk}
\affiliation{Department of Computer Science, University of Bath, Claverton Down, Bath, BA2 7AY, United Kingdom}

\date{\today}

\begin{abstract}

  Rapid advances in AI have opened a new era for physics, enabling unprecedented capabilities in research and discovery. However, these
  advances also bring challenges in verifying, trusting, and reusing the results they so rapidly produce. Formalization 
  is a necessary part of the solution to these challenges.
  In this paper, as well as giving an overview of the role of AI and formalization in physics, we make the case that:  (1) formalization is an essential layer
  for AI-driven discoveries in both computational and theoretical physics; (2) AI is lowering the barrier to, and increasing the efficiency of,
  formalization in physics, though it remains weaker at choosing definitions and designing libraries than at writing proofs; (3) the speed AI brings to formalization in physics currently comes at the cost of trust, credit, and community; (4) shared, community-reviewed formal libraries such as Physlib should be central in making progress in this field, both as a trusted, reusable foundation for physics results and as the environment in which AI systems can reason about physics; (5) formalizing physics raises challenges beyond those in mathematics, most importantly that a checked proof does not establish that its definitions faithfully capture the intended physics. We also give our perspectives on near-term and long-term directions in these areas.

\end{abstract}

\maketitle

\section{Introduction}
\label{sec:intro}

Mathematics provides a very effective language for describing the physical world and deriving predictions that can be compared with observations~\cite{unreasonable-effectiveness}.
Physical reasoning combines mathematical structures with assumptions, approximations, and the specific regime of validity.
Formalization offers a way to make the mathematical content of this reasoning explicit, to check its logical steps, and to organize the results for subsequent use.
There are three closely connected motivations that make formalization of physics appealing: 
(i) verifying physical results, 
(ii) turning individual research contributions into reusable scientific knowledge, and
(iii) establishing general properties across large spaces of models to drive and constrain theory exploration.

Physical arguments have traditionally been communicated through pen and paper or through their digital equivalent, a word processor.
Although digital documents make results easier to share, the reasoning expressed through the encoded mathematical symbols is checked only by hand before being written down, and there is no guarantee of its correctness.
Interactive theorem provers offer a different way to record this knowledge. 
Here definitions, statements, and proofs are written (or formalized) in a formal language, and checking a proof becomes a process analogous to compiling a computer program~\cite{curry1934functionality,curryFeys1958,Howard1969TheFN,Wadler2015PropositionsAsTypes}, which can be fully automated.
The system checks whether the proof establishes the stated conclusion from the given definitions and hypotheses.
A scientific argument can thereby be accompanied by a mechanically checkable account of its mathematical reasoning.

An immediate motivation to formalize physics is to verify the results, from intermediate steps in everyday calculations to the demanding mathematical arguments required to establish a result such as the Yang--Mills mass gap.
Formalization requires assumptions to be stated explicitly and their applicability to be established wherever they are used.
This process can itself sharpen physical understanding by revealing conditions or gaps that remain implicit in an informal derivation.
Its value extends to new results as well as to existing calculations or analysis pipelines, wherever an unnoticed error or an unjustified assumption can propagate into subsequent work.
The need arises both in peer review and in everyday research, when researchers check their own calculations, assess a collaborator's derivation, or supervise a student's work.
Formalization has already begun to find some errors. 
For example, a stability condition of the two-Higgs-doublet potential, from a widely cited 2006 paper~\cite{Maniatis:2006fs}, turned out to miss edge cases, making 
it insufficient in a marginal case~\cite{toobysmith2026formalizingstabilityhiggsdoublet}.

Moreover, the need for verification becomes particularly pressing as AI systems accelerate the production of proposed derivations~\cite{openaiNavierStokes2026, Parisi_2026}. 
An argument generated in minutes may require days of expert scrutiny.
Formalization provides a basis for trusting the mathematical reasoning in AI-generated results: when the intended claim is faithfully formalized and its proof checked, confidence in that reasoning need not depend on the reliability of the AI system that proposed it.
Mechanically checking formalized reasoning reduces the burden of manual verification and allows physicists to devote more attention to physical interpretation, exploration, and creativity.

A further, longer-term motivation is to establish results that specify the properties of large spaces of candidate physical models.
Sampling these spaces and generating examples can reveal promising possibilities and suggest general patterns, but cannot by themselves establish completeness or necessity.
For instance, constructing examples does not establish that all possibilities have been covered, nor does failing to find a counterexample establish that a proposed law always holds.
Ambitious targets include identifying inevitable low-energy signatures of string theory based on clearly specified, broadly applicable assumptions; 
establishing no-go or no-lose theorems for new physics in particle physics and cosmology; 
classifying dark-matter or dark-energy explanations within a defined theoretical framework; 
and characterizing all phases or low-energy behaviors within a specified class of quantum many-body systems.
These are questions of classification, completeness, and necessity, and addressing them requires proofs that control the full class under consideration.
Formalization supports this program by making the scope and assumptions of each statement explicit and enabling the many mathematical results on which it depends to be assembled into a precise and automatically verifiable argument.

A third motivation is to make research results durable and reusable.
Physics produces many valuable calculations and case studies whose intermediate results remain embedded in individual papers, notebooks, or specialized implementations.
The situation resembles scientific software, where a prototype developed to answer one research question is often reused without ever becoming a maintained tool that others can readily apply.
Formalization offers a route from such individual contributions to shared scientific infrastructure.
Definitions, intermediate lemmas, and theorems can become components of a library, with explicit assumptions and interfaces through which subsequent work can use them.
As with scientific software, this requires deliberate attention to design, documentation, review, and maintenance.
In other words, formalizing a result alone does not automatically make it reusable.
The benefit is both immediate and cumulative, because it preserves the value of existing individual calculations and builds the foundations needed for increasingly ambitious formal proofs.

This more structured organization of scientific knowledge must also accommodate the different ways in which physicists describe the same system or concept.
Different formulations can make different properties apparent, support different approximations, or provide more convenient starting points for particular calculations.
Establishing their equivalence under specified assumptions, or identifying precisely where they differ, is itself a source of physical insight.
Since different algorithms may be used to calculate or approximate a physical quantity, formalization can also establish under which assumptions they give equivalent results or, for approximate methods, provide bounds on their differences.
Formal libraries could make these connections explicit, allowing results established in one formulation to be used in another once the necessary correspondence has been proved.

So far, the overhead of formalization has made it difficult to justify within ordinary research practice for many physicists.
Learning a proof assistant requires substantial effort, and even an experienced user may need considerable time to supply details that a conventional derivation leaves implicit.
Where expert checking appeared sufficient and the reusable benefits were uncertain or distant, this additional work could reasonably seem disproportionate to its immediate scientific return.
As coding agents take on more of the work of writing and revising formal code in response to feedback from the theorem prover~\cite{liu2026diveclaudecodedesign,gen-ai-models}, this cost is now rapidly decreasing.
This changes the practical case for adoption. 
Projects for which formalization previously appeared prohibitively expensive become more accessible.
However, expert judgment remains necessary to choose appropriate definitions, assess their physical meaning, and organize the resulting code for reuse.

The developments reviewed below provide starting points for this transition.
Selected physical definitions, calculations, and research results have been formalized, drawing on mathematical foundations in Mathlib~\cite{mathlib,lean} and contributing to physics libraries such as Physlib~\cite{PhysLib,heplean,Meiburg:2025mwn}.
These achievements demonstrate the feasibility of checking individual results and building reusable foundations~\cite{perspective}.
Making these capabilities part of routine physics research requires extending their coverage and connecting them systematically. 
Researchers need to be able to express a new problem using existing definitions and results without rebuilding its foundations each time.
The longer-term classification goals require these foundations to support statements across entire specified classes of models.
The program therefore involves both formalizing further results and developing the shared infrastructure through which those results can be combined.

In this perspective, we examine how the interaction between AI and formalization could advance in the future. 
We consider both how AI can assist the formalization of physics and how, in the longer run, formalized knowledge can support more reliable AI reasoning and accelerate scientific discovery.
We assess present capabilities as well as near-term and longer-term opportunities.
After the brief overview of formal methods below, Section~\ref{sec:AIforForm} examines how AI can advance formalization in physics. 
We review current capabilities, their implications for the trustworthiness, quality, and sustainability of formalization projects, and opportunities ranging from improved planning and review to changes in how physics is communicated and taught.
Section~\ref{sec:FormforAI} considers the reverse direction. 
It looks at how formalization can support AI applications in physics, from checking and providing feedback on AI-assisted derivations and computational methods to enabling scientific discovery.
We discuss the role of shared formal libraries and the limitations of formal verification in assessing physical validity.
The Conclusion considers the expertise, infrastructure, and academic incentives needed for the systematic integration of formal methods and AI, which could accelerate scientific progress in physics in the coming years.

\subsection{Formalization and interactive theorem provers}
Formalization is the process of expressing definitions, assumptions, statements, and proofs in a precise formal language that a proof assistant (also called an interactive theorem prover) can mechanically check.
This involves translating their informal presentation into a structured and unambiguous system that is executable as code within an interactive theorem prover.
In physics, the starting point typically combines natural-language explanations, mathematical expressions, and implicit conventions.
The translation requires making the mathematical objects and assumptions explicit and supplying the logical steps needed to establish the stated conclusions.

An interactive theorem prover acts as a proof compiler. The user supplies definitions, statements, and proofs, and the system checks that the proofs establish the stated conclusions from the specified hypotheses within its logical foundations.
This is closely related to software verification, where one proves that a program satisfies a mathematical specification.
In physics, the statement to be checked may instead concern a physical model, a derivation, or an algorithm for calculating an observable.
In each case, the guarantee concerns the formal statement; whether that statement captures the intended scientific or computational claim requires separate assessment.
More specifically, it is essential to distinguish the verification of mathematical reasoning from the validation of its physical meaning.
A checked proof establishes a conclusion from formal definitions and hypotheses.
Assessing whether those definitions express the intended physics, whether the assumptions and approximations are appropriate, and whether the model describes observations requires additional scientific judgment and empirical evidence.
Formalization supports this assessment by exposing assumptions and dependencies, but the physical value of a formal result depends on the connection between the checked statement and the scientific claim it is intended to express.

In this perspective, we use Lean~\cite{lean} as our reference interactive theorem prover and as the language for our examples.
Its integration with mathematical libraries such as Mathlib~\cite{mathlib} and physics libraries such as Physlib~\cite{PhysLib} makes it a useful setting for discussing the formalization of physics, especially given its popularity with current AI applications.
Other proof assistants, including Agda, Isabelle, and Rocq, also support formal reasoning.
Although their foundations and implementations differ, many of the opportunities and challenges discussed here extend beyond Lean.

Lean is based on dependent type theory~\cite{AvigadMassot2020,nederpelt_geuvers_2014}.
A type describes a kind of mathematical object, and an object of that type is called a term.
The notation $a : T$ states that the term $a$ has type $T$; for example, $m : \mathbb{R}$ declares that $m$ is a real number.
Types can also describe structured objects that combine several pieces of data with proofs of conditions on those data.
In the particle example below, we will combine a mass and a velocity with a proof that the mass is positive.
This allows assumptions to be carried explicitly by the objects to which they apply.

The connection between types and proofs is expressed by the Curry--Howard correspondence: propositions can be represented as types, and proofs as terms of those types~\cite{Howard1969TheFN}.
If $P$ is a proposition, constructing a term $h : P$ amounts to supplying a proof of $P$.
For example, a term of type $0 < m$ is a proof that $m$ is positive.
A proof of an implication $P \to Q$ can be understood as a function that takes a proof of $P$ and produces a proof of $Q$.
Lean checks that these constructions have the required types, thereby checking the corresponding mathematical reasoning.

Writing proof terms directly can be cumbersome.
Interactive theorem provers therefore provide tactics. These are commands that help construct a proof by applying known results, simplifying expressions, or breaking a statement into smaller proof obligations.
During this process, Lean displays the available hypotheses and the remaining goals.
The tactics ultimately produce a proof term that Lean's kernel checks.
This separates the task of finding a proof, which may involve a person, an automated tactic, or an AI system, from checking the resulting proof.

A small example illustrates how data and assumptions are combined.
We consider a nonrelativistic particle moving in one spatial dimension, represented here only by its mass and velocity.
The declaration \myinline{mass_pos : 0 < mass} introduces a field named \myinline{mass_pos} whose value is a proof that the mass is strictly positive.
Constructing an instance of \myinline{Particle} therefore requires supplying both the numerical data and a proof of this condition.
This deliberately minimal model does not describe charge, spin, helicity, or other particle properties; its scope and physical interpretation should be documented alongside the formal definition.
We define its kinetic energy by the usual expression $E_{\mathrm{kin}}=\tfrac12 mv^2$.

\begin{code}
/-- A nonrelativistic particle in one dimension,
    with positive mass; other properties omitted. -/
structure Particle where
  mass : ℝ
  velocity : ℝ
  mass_pos : 0 < mass

noncomputable def kineticEnergy (p : Particle) : ℝ :=
  1 / 2 * p.mass * p.velocity ^ 2
\end{code}

We can now state and prove that the kinetic energy is non-negative for every particle described by this structure.
The positive-mass assumption is available through the field \myinline{p.mass_pos}, while the square of the real-valued velocity is non-negative.
In the proof below, the comments display Lean's proof state after each step: the expressions above the symbol $\vdash$ (called the turnstile) are the available objects and hypotheses, and the expression following it is the goal that remains to be proved.
Together, the steps construct a proof of the non-negativity statement.

\begin{code}
theorem kineticEnergy_nonneg (p : Particle) : 0 ≤ kineticEnergy p := by
  -- p : Particle
  -- ⊢ 0 ≤ kineticEnergy p
  unfold kineticEnergy
  -- ⊢ 0 ≤ 1 / 2 * p.mass * p.velocity ^ 2
  have hm := p.mass_pos
  -- hm : 0 < p.mass
  -- ⊢ 0 ≤ 1 / 2 * p.mass * p.velocity ^ 2
  positivity
  -- goals accomplished
\end{code}

The tactic \myinline{unfold} replaces \myinline{kineticEnergy} by its defining expression.
The command \myinline{have hm := p.mass_pos} makes the positive-mass hypothesis available under the name \myinline{hm}.
Finally, \myinline{positivity} establishes that the product is non-negative using the positivity of the numerical factor and the mass, together with the non-negativity of the squared velocity.
These steps produce a proof term that Lean's kernel checks against the stated theorem.
Once established, this theorem can be reused in subsequent proofs without repeating its derivation.

\begin{figure*}
  \centering
  \begin{tikzpicture}[
      level/.style={
        draw=ladderInk, line width=0.8pt, rounded corners=1.5pt, fill=white,
        text width=3.3cm, minimum height=1.7cm, align=center, text=ladderInk
      },
      spine/.style={line width=1.4pt},
      rung/.style={line width=0.8pt},
      label/.style={anchor=north, align=flush left, text=ladderInk, inner sep=0pt, text width=4.3cm},
    ]

    \node[level] (proof) at (13.65, 0)
      {\textbf{Proofs}\\[1pt]{\footnotesize\color{ladderSubtle}type checking, tactics}};
    \node[level] (stmt) at (9.1, 0)
      {\textbf{Definitions and statements}\\[1pt]{\footnotesize\color{ladderSubtle}the mathematics, in Lean}};
    \node[level] (model) at (4.55, 0)
      {\textbf{Physical model}\\[1pt]{\footnotesize\color{ladderSubtle}assumptions, regime}};
    \node[level] (world) at (0, 0)
      {\textbf{The world}\\[1pt]{\footnotesize\color{ladderSubtle}experiments, observations}};

    \newcommand{\ladderlink}[7]{%
      \coordinate (m) at ($(#1.east)!0.5!(#2.west)$);
      \draw[spine, #3, -{Stealth[length=6pt,width=5pt]}] (#1.east) -- (#2.west);
      \fill[#3] (m) circle (2.4pt);
      \draw[rung, #3] (m) -- ++(0,-1.25) coordinate (e);
      \node[anchor=north, inner sep=0pt, text=#3, font=\bfseries] (t) at ($(e)+(0,-0.15)$) {#4};
      \node[label] at ($(t.south)+(0,-0.2)$)
        {{\small #5}\\[1pt]
         {\footnotesize\itshape\color{ladderSubtle}#6}\\[1pt]
         {\footnotesize\emph{Example:} #7}};
    }

    \ladderlink{stmt}{proof}{ladderVerif}{Verification}
      {Is the mathematics consistent, and does the proof establish the stated result?}
      {Checked by the proof compiler: everything type-checks, with no \texttt{sorry} and no unintended axioms.}
      {Lean accepts \texttt{kineticEnergy\_nonneg}, so $\tfrac12 mv^2 \ge 0$ for every \texttt{Particle}.}

    \ladderlink{model}{stmt}{ladderValid}{Logical validation}
      {Do the definitions and statement faithfully express the physical model?}
      {Checked by expert review of definitions and hypotheses.}
      {Is \texttt{Particle}, with $E_{\mathrm{kin}}=\tfrac12 mv^2$, a faithful description of a nonrelativistic particle?}

    \ladderlink{world}{model}{ladderExp}{Experimental validation}
      {Is the physical model correct within its stated regime?}
      {Checked by experiment and observation.}
      {In the classical regime, $v \ll c$, experiments confirm that the kinetic energy is $\tfrac12 mv^2$.}
  \end{tikzpicture}
  \caption{From the physical world to Lean code, illustrated with the kinetic-energy example of this section.
    The theorem prover guarantees only the final step, verification.
    Whether the formal definitions capture the intended physical model (logical validation), and whether that model describes nature (experimental validation), require scientific judgment and empirical evidence.}
  \label{fig:verification-ladder}
\end{figure*}
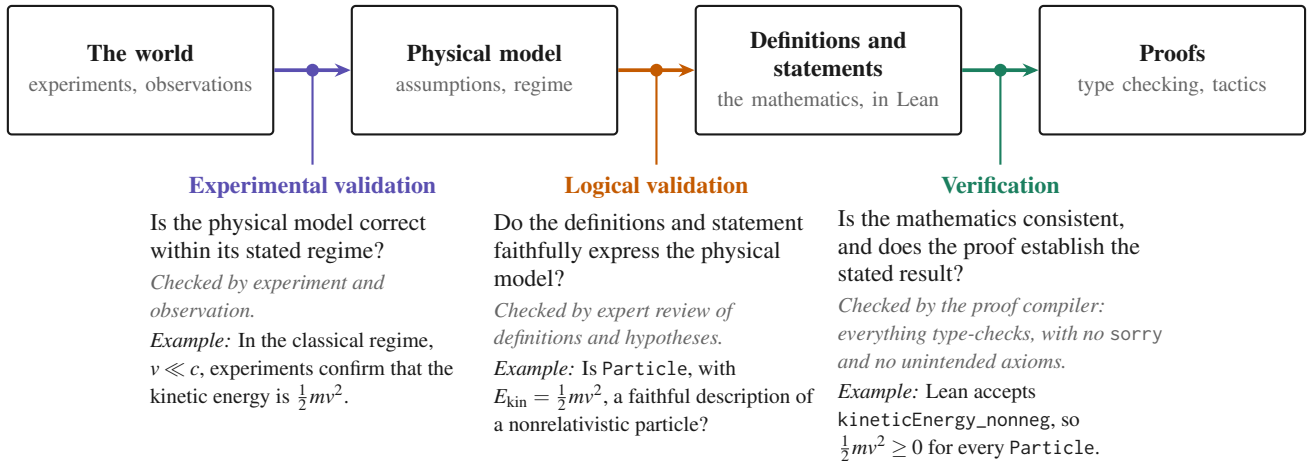

The checked theorem establishes the non-negativity of the kinetic energy within the model we defined.
It does not establish a definition of a particle suitable for all circumstances. 
The structure is useful for this calculation, while omitting properties needed in other settings.
Its physical interpretation, assumptions, and intended scope should therefore be clearly documented and assessed through scientific review.
Formalizing physics requires a clear connection between mathematical objects and their physical interpretation.
As shown in Fig.~\ref{fig:verification-ladder}, three questions must be distinguished: 
does the proof establish the formal statement, 
does that statement faithfully express the intended physical argument, and 
are the assumptions appropriate for the system being studied.
An interactive theorem prover addresses only the first question.
The others require scientific judgment, supported by documentation, comparison with established results, and, where relevant, observational or experimental evidence.
Documentation should therefore explain what the formal objects represent, which assumptions are being made, and the intended regime of validity.
We return to this asymmetry between checking a proof and validating its definitions when discussing trust in Section~\ref{sec:AIforForm}. 

\subsection{Formalization efforts}
\paragraph{Mathematics}
The formalization of mathematics has a history extending well beyond the recent advances in AI.
Early systems such as Automath established ways of expressing mathematical reasoning in a computer-checkable language~\cite{automath1967}.
Subsequent milestones include the formal proofs of the four-color theorem~\cite{gonthierFourColour}, the Kepler conjecture~\cite{halesKepler2015}, and the Liquid Tensor Experiment~\cite{liquidTensor2022}, which formalized a substantial result in contemporary mathematics.
Together, these projects demonstrated that formal verification can address both proofs involving extensive computation and sophisticated mathematical theories.

A central foundation for mathematical formalization in Lean is Mathlib, a community-developed library of definitions and theorems spanning many areas of mathematics~\cite{mathlib}.
The project's statistics page, consulted on 24 September 2026, reports 137,019 definitions, 289,120 theorems, and 772 contributors~\cite{mathlibStats}.
Its shared infrastructure allows new projects to build on existing results rather than reconstruct their mathematical foundations.
Contributions undergo review for generality, integration with the library, and maintainability, alongside automatic proof checking~\cite{mathlibReview}.
These complementary requirements support both the correctness of individual results and their usefulness in subsequent formalizations by humans or AI systems.

Large formalization projects combine mathematical expertise with the development of reusable definitions and intermediate results.
A blueprint can organize this work by setting out the mathematical argument and the dependencies between its component statements~\cite{leanblueprint}.
A prominent example is the project led by Kevin Buzzard to formalize Fermat's Last Theorem, following a mathematical route developed with Richard Taylor~\cite{fltProject}.
Such projects illustrate how a major target theorem can guide the coordinated development of the substantial mathematical infrastructure needed to prove it.

AI-driven formalization is now being pursued both through shared libraries and through projects targeting particular theorems.
Tau Ceti is a library built on Mathlib in which AI contributors implement mathematics guided by human-written roadmaps and review criteria~\cite{taucetiproject2026}.
In September 2026, Anthropic released a complete Lean formalization of Fermat's Last Theorem, building in part on the existing community project~\cite{anthropicFLT2026}.
OpenAI also released an analytical proof of finite-time blowup for the forced Navier--Stokes equations, accompanied by a Lean formalization~\cite{openaiNavierStokes2026}.
These efforts illustrate different roles for AI. They include extending reusable mathematical infrastructure, formalizing established proofs, and supplying formal verification alongside new mathematical results.

\paragraph{Physics}
Formalization in physics has a history spanning more than a decade, with projects addressing different scientific needs~\cite{perspective}.
Work on optical systems has used theorem proving to verify properties of physical models relevant to engineering applications~\cite{khan2014formal}, with related work formalizing quantum optics~\cite{vermeulenformal}.
Other projects have formalized elements of special relativity in Rocq~\cite{lu2017formalization} and derivations in chemical physics and thermodynamics in Lean~\cite{josephsonThermo}, making their assumptions and mathematical dependencies explicit.
In quantum information theory, the formalization of the generalized quantum Stein's lemma checked a corrected proof following the discovery of a gap in the original argument, and identified further minor imprecisions~\cite{Meiburg:2025mwn}.
These examples illustrate motivations ranging from reliable system analysis and clarification of established derivations to verification of research-level results.

Physlib aims to provide a shared library for physics analogous to the role of Mathlib in mathematics~\cite{PhysLib}.
Formed by bringing together PhysLean and Lean-QuantumInfo, it develops formal definitions, theorems, and calculations for use across physics~\cite{heplean,meiburg2024quantuminfo}.
Its coverage includes basic material in areas such as electromagnetism, alongside developments in string theory, quantum field theory, and quantum information.
The project accommodates different physical approaches and starting assumptions without requiring a single foundational axiomatization of physics.
Its core library emphasizes review, physical documentation, readability, and reuse.
A separate development layer, PhyslibAlpha, supports contributions under a lighter review process, making the distinction between stages of curation explicit~\cite{PhysLib}.

AI-assisted formalization is also being applied to substantial arguments in mathematical physics.
For example, Douglas et al.~\cite{douglas2026formalizationqft} formalized the construction of a free massive scalar field in four-dimensional Euclidean spacetime and proved that it satisfies the Glimm--Jaffe axioms, a variant of the Osterwalder--Schrader axioms.
Their work combined expert choices of definitions and mathematical strategy with coding assistants, including Claude Code, and Lean-specific tooling.
Mathematical results initially left as assumptions were subsequently proved or connected to Lean's mathematical foundations and Mathlib.

These efforts are accompanied by the development of specialized agents and tools for formalization in mathematics and physics.
We discuss their capabilities and limitations in Section~\ref{sec:AIforForm}.

\begin{figure*}
  \centering
  \includegraphics[width=\textwidth]{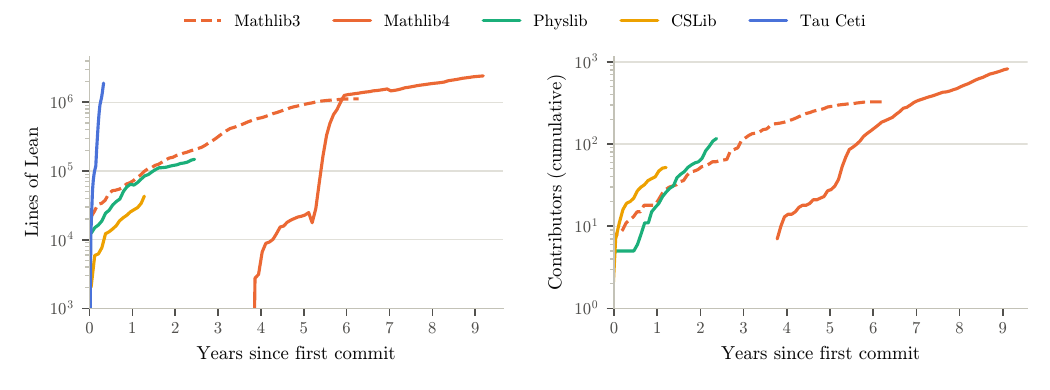}
  \caption{Growth of Mathlib3 (dashed) and its successor Mathlib4, Physlib, CSLib, and Tau Ceti, plotted against the years since each library's first commit.
    Mathlib4 is measured from the first commit of Mathlib3, so that the two appear in their historical order; every other library is measured from its own first commit.
    \textbf{Left}: total lines of Lean code on the default branch; for Physlib only the core library is counted, excluding PhyslibAlpha and QuantumInfo, which live in the same repository.
    \textbf{Right}: cumulative number of distinct contributors (bots excluded).
    Both are shown on a logarithmic scale.
    Tau Ceti is omitted from the right panel since its code is mainly written by AI agents.
    Data computed from the git history of each library, as of 30 September 2026.}
  \label{fig:library-growth}
\end{figure*}

\paragraph{Other applications}
Formal methods have long been used in software and hardware verification to establish that implementations satisfy specified properties~\cite{project-everest}.
Examples include CompCert's verification of compiler correctness~\cite{compcert} and the use of ACL2 to verify floating-point arithmetic in commercial processors~\cite{acl2Applications}.
These applications demonstrate that machine-checked reasoning can support practical engineering systems.
More broadly, CSLib aims to provide shared infrastructure in Lean for computer-science theorems and verified programs, analogous to the role of Mathlib in mathematics~\cite{barrett2026cslibleancomputerscience}. The growth of Mathlib, Physlib, CSLib, and Tau Ceti is shown in Fig.~\ref{fig:library-growth}.

Formalization is also being explored in cryptography, economics, finance, chemistry, biology, and scientific computing.
Examples include the verification of both the theoretical derivations and the implementations of cryptographic algorithms, EconCSLib's formalization of research in economics and computation~\cite{garg2026econcslib}, libraries for mathematical finance~\cite{coelho2026finance}, and the chemical theories discussed above~\cite{josephsonThermo,ugwuanyi2026leanbetformallyverifiedsurfacearea}.
In mathematical biology, work on the continuous $\pi$-calculus formalizes a language for describing biochemical interactions~\cite{leanCpi}.
Scientific-computing projects such as SciLean explore how formalized mathematics can guide the construction and transformation of numerical programs~\cite{scilean}.
Among other applications, some early-stage companies are also exploring formalization for regulation~\cite{pramana}.
These efforts range from individual demonstrations to libraries under active development, with differing levels of coverage and maturity.

Across these applications, formalization makes definitions and assumptions explicit and checks the conclusions drawn from them.
Its usefulness depends on identifying which parts of an argument admit a precise mathematical or logical formulation, and how those parts connect to observations, measurements, or other evidence.
Formal proofs can establish consequences of a model, while assessing the model's relevance requires knowledge of the domain.
In physics, this connection raises particular questions about physical interpretation, approximations, and the choice of mathematical description, which we discuss next.

\subsection{Physics formalization challenges}
Validation requires particular care and does not offer the same guarantees as theorem proving. As discussed previously, this is particularly challenging in physics compared to mathematics, given the greater ambiguity of its concepts and the need to relate the physical model and its mathematics to empirical observations.
Humans can review the formalization with literature support, perhaps aggregated and summarized by an AI model. 
Another validation approach is to write many different Lean lemmas that employ the target definition, making it increasingly unlikely that they are all satisfied by coincidence. Finally, definitions can be characterized: they can be pinned down by their crucial relations and properties, known special cases, symmetries, and limits, which then serve as the equivalent of unit tests supporting the validity of the definition. In many physics projects, the acceptance of a proposed definition requires that characterizing lemmas have been proven.

Physical concepts often admit several mathematical descriptions, each suited to different questions or regimes.
Electromagnetism, for example, can be described using fields and potentials, with different mathematical requirements depending on whether these are treated as smooth functions, distributions, or geometric objects such as connections.
The choice of mathematical definitions affects which properties are explicit, which results are readily expressible, and how the formalization connects to existing libraries.
Identifying suitable definitions and establishing relationships between alternative descriptions is therefore a substantial part of formalizing physics, and can itself clarify the underlying theory.

Physics papers often combine definitions, assumptions, calculations, and physical interpretation without separating them into explicit mathematical statements.
Intermediate steps may be omitted, conventions left implicit, and approximations introduced without a complete account of their conditions of applicability.
Formalization requires reconstructing this structure. 
In particular, it involves identifying the objects under discussion, stating the hypotheses and conclusions, and supplying the mathematical steps that connect them.
Where an argument uses an approximation, the formal statement must make clear what is being approximated and what claim is established within that approximation.
This process can expose missing conditions and distinguish an established consequence from a heuristic expectation.

Formalizing physics does not require first constructing a single axiomatic foundation for the whole subject.
A physics formalization can choose where its argument begins.
One approach is to work within an established axiomatic framework and derive consequences from its assumptions~\cite{douglas2026formalizationqft,lu2017formalization}.
Another is to start from the model, equations, or physical assumptions used in a particular calculation and formalize the reasoning from that point onwards.
This establishes a conditional result: given the specified starting assumptions, the conclusion follows.
Such a result can be useful even when a derivation of the starting model from a more fundamental theory is unavailable.

The appropriate starting point depends on the scientific question.
Assumptions chosen too narrowly may limit the usefulness of a result, while starting from very general foundations may require substantial work before the calculation of interest becomes accessible.
The aim is to choose assumptions that make the intended claim meaningful and its proof manageable, while retaining sufficient generality for subsequent use.
The scope of the resulting theorem should be explicit, so that others can assess when its assumptions apply.
Later, these results can be connected to a more fundamental or alternative description.
One can establish these connections by proving that another model satisfies the required assumptions, or by deriving the relevant equations in a specified limit or approximation.
This is the approach taken in Physlib~\cite{PhysLib}, where individual developments can be useful on their own while also providing opportunities to establish relationships between them.

Physical assumptions can be introduced in two ways.
Encoding them as theorem hypotheses or conditions on a structure makes the resulting claims explicitly conditional.
Introducing them through Lean's \myinline{axiom} command instead enlarges the set of statements accepted without proof. This is discouraged because incompatible additions can compromise the logical foundations of subsequent results.
The former approach makes the dependence on a physical model explicit and supports reuse across models satisfying the same conditions.
It also separates proving consequences of those conditions from establishing that a model satisfying them exists.

\subsection{Formalization workflows}
\label{subsec:intro:formalization-workflows}
The effort required to formalize a physical result depends strongly on the definitions and theorems already available.
Even a calculation using familiar mathematics may require substantial preparation if its physical concepts, conventions, or intermediate results have not yet been formalized.
Building these foundations in a reusable form allows later projects to benefit from that work.
Progress therefore involves both verifying particular results and developing shared libraries that make further formalization easier.
This subsection discusses how these two objectives can guide the organization of a formalization project.
There are two complementary approaches: building a library around the underlying definitions, and starting from the particular results to be formalized.
Both approaches apply in mathematics and physics, as well as across other disciplines.

The first approach starts from the underlying definitions. 
One identifies the definitions on which the field is based, connects them to existing mathematical infrastructure such as Mathlib, and then proves the associated properties and results.
A more general library is built around the topic to be formalized, checking not only that each object can be used in the required theorems, but also that it supports other theorems expected to hold.
This approach requires a far-sighted perspective, because it is necessary to examine the objects and definitions on which a specific field depends.
This is the ``library-building'' approach.

The second approach is to start from the target results to be formalized and work backwards through their dependencies until they can be connected to Mathlib, Physlib or another existing library.
In this way, it is possible to lay out the main results of a field before all the intermediate results have been implemented, although statements with unresolved proof obligations remain unverified.
Depending on the topic, it may be hard to immediately identify the right way to express a physics result formally and to adopt definitions that will be useful beyond the immediate application.
There are often multiple ways to mathematically define the fundamental objects of a theory, and an unsuitable choice may result in much harder formalization work and the need for a future redesign of the results depending on it.

In practice, one must exercise judgment in combining these approaches, deciding how much general infrastructure to develop alongside a particular result.
Both require the same standard of proof, but differ in how the work is organized.
As the library grows, existing definitions and theorems can be reused, reducing the work required to formalize further results.

\section{AI to advance physics formalization}
\label{sec:AIforForm}
In this section, we focus on the use of artificial intelligence to support and accelerate physics formalization.
General language models are useful for retrieving and discussing scientific knowledge, identifying how a physical concept is expressed mathematically, and finding relevant references.
Code assistants can lower the access barrier to formal methods by helping at the syntax level, exploring existing formal libraries, and proposing code to implement a requested theorem or definition.

These tools can integrate directly with the development environment and GitHub, including creating or commenting on pull requests.
Agentic coding assistants can write and implement longer sections of code autonomously, using feedback from the compiler to revise their output, thereby reducing the work required from the developer.

However, as we will discuss in the following subsections, these innovations also come with several challenges, such as code review, validation and trust concerns.

\subsection{State-of-the-art AI for formalization}
We begin by reviewing the performance and effectiveness of AI-assisted formalization.
This field is evolving extremely rapidly: it has changed dramatically since mid-2025, and it will probably change as dramatically again in 2027.

The formalization process can be split into two parts: first, writing formal statements of definitions, properties, and theorems; second, proving these statements.
We start by analyzing these two components (in reverse order), and subsequently examine the workflows in which AI tries to iteratively and autonomously formalize more than a single statement or theorem at once.

\paragraph{Theorem proving}
The most straightforward automation that AI enables is theorem proving. 
Since the compiler can check a given proof and provide feedback, AI systems can autonomously explore the proof space (which itself might require many lemmas) until they find a proof that compiles.
Given a formal theorem statement in Lean, AI models can fill in the verifiable proof, so that the user can focus on getting the right formalization of the objects, and offload the proof to AI systems.
Remarkably, AI output can essentially be trusted in this setting.
If a proof compiles, proves the intended statement, contains no unresolved proof obligations, and relies only on the accepted axioms, we can accept it; the main work becomes improving its style, readability, and perhaps compilation efficiency.
Compiler bugs are a further concern, and additional assurance can be obtained by running the proof through external checkers~\cite{comparator,carneiro2025lean4leanverifyingtypecheckerlean}.

As a consequence, AI has become better at generating proofs within interactive theorem provers.
Following the pioneering work on AlphaProof~\cite{hubert2026olympiad}, several AI proof assistants are now available.
Among them, Harmonic's Aristotle is relatively popular~\cite{achim2025aristotleimolevelautomatedtheorem}, but several open-source initiatives exist as well~\cite{ren2025deepseekproverv2advancingformalmathematical,axiommath2026imo2026}.
Even general-purpose large language models such as Claude or Gemini are now very good at proving results. 
At the moment, the latest general-purpose models~\cite{sienicki2026humanauditopenaisaigenerated} or agentic harnesses seem to be the most successful approaches~\cite{letson2026sorrydbaiproverscomplete}, and tend to outperform one-shot specialized large language models~\cite{chen2025seedprover15masteringundergraduatelevel,axiommath2026imo2026}. 
This has been demonstrated not only with results such as proving Erd\H{o}s problems~\cite{sothanaphan2026resolutionerdhosproblem728, tsoukalas2026advancingmathematicsresearchaidriven}, but also on a more day-to-day basis with more applied formalization projects~\cite{hariharan2026progressformalizingspherepacking} and in physics~\cite{douglas2026formalizationqft, ilin2026semiautonomousformalizationvlasovmaxwelllandauequilibrium}. 
Systematic benchmarks, ranging from International Mathematical Olympiad and undergraduate mathematics problems to physics fields such as quantum information, have been developed to assess progress~\cite{zheng2022minif2fcrosssystembenchmarkformal,tsoukalas2024putnambench,li2026lean4physics,zhang2026benchmarkingquantumtheorems,yin2026axqmtextbookscalebenchmarkformal}.

\paragraph{Statement formalization}
On the other hand, much more trust is required when writing a formal definition, because one cannot unambiguously check whether its structure and design are appropriate and complete with respect to a given informal reference.
The comparison is not objective, and there are many choices to be made when inferring a formal version of a concept described in natural language.
For this reason, it is also hard to verify whether a proposed formalization is acceptable, which makes it harder for AI to get feedback and iterate.
As previously mentioned, directly formalizing one statement at a time, without considering the full context and the broader picture, may produce suboptimal definitions that only partially have the properties of the target~\cite{ammanamanchi2026faultsformalbenchmarkingdataset, feng2026characterizingparaphraseinducedfailureslean}.
For example, one may define a more specific version of the object, or add extra assumptions that make the derivation simpler.
As a consequence, when it comes to definitions, large language models are slightly more brittle, and can lead to embarrassing failures, but they are steadily improving and gaining the trust of practitioners~\cite{ren2026merleanagenticframeworkautoformalization,milikic2026leanflowcasestudyworkflowdriven,ying2025leanworkbooklargescalelean}.

\paragraph{End-to-end workflows}

Finally, one area in which we notice particular weaknesses is library design and the organization of a project to complete larger-scale formalization tasks autonomously.
The field is evolving quickly, but no safe out-of-the-box solution exists yet.
In particular, as described in the next subsection, we cannot rely on current coding assistant models, especially general-purpose ones, to automatically produce well-designed results that fit into a good library, following the same standards as human-written libraries such as Mathlib.
This is due to the current limitations of large language models, which are still affected by shortcomings in long-term planning.

Of the approaches described in Section~\ref{subsec:intro:formalization-workflows}, current agentic approaches and coding assistants tend to formalize statements mostly bottom-up: starting from the core results, and then tracing back all the dependencies, or sequentially, from the beginning to the end, when given a topic from a book or paper.
In contrast, a mathematical or theoretical physicist would often plan very carefully top-down from the fundamental definitions to the intermediate results necessary to implement the target object, or find a good compromise between the two approaches.
End-to-end workflows can also be guided by blueprints or other explicit formalization plans, as discussed in Section~\ref{subsec:AIForm:near-term}.

Several commercial startups and nonprofit organizations are developing specialized, often proprietary, systems to support the end-to-end development of Lean formalizations. 
We identify three main formalization workflows:
\begin{itemize}
  \item human-written, using standard tooling and very limited use of AI,
  \item AI-crafted, where a human drives the high-level design and the model suggests definitions and fills in the proofs interactively,
  \item AI-blitz, where a general model is asked to produce as much as possible with minimal intervention, and the human analyzes it afterwards.
\end{itemize}

Today, end-to-end automated formalizations can still be hard for humans to read, 
and fall short of the standards expected in human-written formal libraries.
This has created a separation in the ecosystem between libraries, or 
parts of libraries. For example, Tau Ceti~\cite{taucetiproject2026} for mathematics and PhyslibAlpha for physics are designed to scale more effectively with AI-assisted formalizations, whereas traditional libraries like Mathlib and Physlib maintain standards that often require careful human oversight and 
will often close low-quality AI-generated pull requests~\cite{mathlibAIContributions}. Tau Ceti is overcoming some of these challenges with a careful AI-led review cycle. 

That said, these end-to-end formalizations can be useful for setting a benchmark and establishing a foundation for what can be done. 
Some notable examples include ``130k Lines of Formal Topology in Two Weeks''~\cite{urban2026130klinesformaltopology} and the formalization of QFT~\cite{douglas2026formalizationqft}. 
In addition, there have been further attempts in physics~\cite{meadows2026formalscience,yin2026axqmtextbookscalebenchmarkformal,ren2026merleanagenticframeworkautoformalization}.

\subsection{Dimensions of formalization and AI}
\label{subsec:AIForm:dimensions}

Considered more broadly, formalization involves much more than simply taking a result and proving it within a language like Lean. 
In our opinion, there are six dimensions worth discussing, relating to formalization: trust, readability, quality, efficiency, community
and sustainability.
There is a seventh dimension, applicability, that is also worth discussing, but this is slightly different from the rest.

As discussed before, there are some general strengths and weaknesses of using AI for formalization across these dimensions.
The first strength is that AI is obviously very fast.
With AI, it is very easy to write new code and to modify existing code. 
It also lowers the barrier to entry, allowing more people to engage in formalization than before and spreading its benefits more widely. 
AI is also strong at tasks which are mechanical, local, and checkable. 
In Lean, an example of such a task is proving a theorem that has already been stated, or keeping documentation in sync with the code beneath it.
The main weakness is that whenever a task requires more judgment, for example the end-to-end formalizations discussed above or the generation of whole libraries, the results tend to be weaker. 
Results end up in the wrong place, and lemmas are proved that already exist elsewhere in the library. 

\paragraph{Trust}
Trust is what lets somebody other than the author rely on a result without re-deriving or fully checking it themselves.
Checking a proof in Lean guarantees that the stated conclusion follows from the given definitions and assumptions, but that guarantee does not cover everything.
For example, Lean does not track the physics context, so a theorem may be mathematically correct yet be the wrong theorem, and Lean cannot tell whether a definition is the one that was intended. 
Definitions must simply be trusted, and anybody relying on a formalization has to go back and check them. 
We have seen this in Section~\ref{sec:intro}, and it corresponds to the difference between validation and verification.
In the example in Fig.~\ref{fig:verification-ladder}, Lean checks that $0 \leq E_{\mathrm{kin}}$ follows from the definitions and assumptions. Whether these describe the intended nonrelativistic particle requires scientific judgment; whether the model applies to the physical system requires empirical evidence.

Projects can therefore build trust in their definitions in two ways:
\begin{itemize}
  \item trust by authority, where a review process ensures somebody other than the author has checked the content, bringing peer review into Lean;
  \item trust by reuse, where the downstream uses of a definition, and failure modes set up around it, provide evidence that it is the right one.
\end{itemize}

The main challenge of AI adoption comes from its speed. 
The tendency to formalize as much as possible as quickly as possible produces massive repositories of Lean code, almost none of which has been checked by a human, removing trust by authority entirely. 
If no human has ever checked the definitions, it is debatable whether the results can be trusted at all.
An LLM reviewer can mitigate this problem, but it does not fully remove it. 
One then has to trust the AI reviewer instead, which is itself a personal decision.

However, we can use AI to strengthen trust by reuse, because it is good at generating the sanity checks and failure modes that would fail if a definition were wrong.
AI can also take on the mechanical part of review, freeing the human reviewer for the parts requiring judgment, or it can help an author present a contribution so that the reviewer's time is spent judging the most relevant parts of the work rather than reconstructing or decoding it.

\paragraph{Readability}
Trust by authority is possible only if the human reviewer can follow the code. 
This works if the code itself is readable. 
In Lean, this means following conventions, keeping documentation concise, and organizing files logically so that results can be found intuitively. 
Therefore, while valid large-scale formalization efforts can be useful for AI consumption, for example as training data, benchmarks or reference corpora, those meant as reusable components that form the foundations of a formal library that humans should trust require a proper structure, appropriate concept naming, extensive explanatory comments, and a well-designed proof style.
For example, large-scale efforts such as Anthropic's recent formalization of Fermat's Last Theorem are correct, but can be much longer than necessary~\cite{anthropicFLT2026}.
It would be challenging for a human to review 13 million lines of code.

A formal language does not check style. 
Unreadable code compiles just as well as readable code, and nothing within the system flags code as unreadable or unreviewable. 

AI helps with readability in the sense that it is very good at correcting documentation when the source code has changed underneath it, or at checking whether the documentation still matches the code. 
It can also be used to summarize files and guide newcomers through projects. 
On the other hand, when it comes to producing readable code, AI is not yet good enough. 
This is likely because it requires whole-project judgment and an understanding of what makes something readable to humans and what does not. 
Anecdotally, it is still very easy to distinguish AI-written code in Lean from human-written code because of this. 

\paragraph{Quality}
Quality extends readability. While readability asks whether a human can follow the code as it stands, quality asks what happens to that code afterwards. 
For example, it asks whether the results are reusable, whether the code can be maintained as the project changes around it, and whether it sits at the right level of generality to be used by others.

Results that do not reach a certain level of quality will often need to be restated, no matter how readable the code is. 
Much of the reason why projects like Mathlib succeed is that they have very high-quality code. 
This is ensured by a strict review process.
The library is easy to search, and its code is at the right level of generality, making it reusable across different projects. 

AI helps most clearly with the quality of code that already exists, since golfing a proof, or finding results buried inside existing proofs that could be pulled out into independent reusable lemmas, is local and checkable.
These are improvements a project would like to make continuously, but that nobody has the time to make by hand.

While frontier models are now producing proofs of acceptable quality, AI remains noticeably weaker at writing good-quality libraries. 
This is because the judgments that quality depends on are global.
These require holding the shape of the whole project in mind, which a model working on one statement at a time does not do.

As a matter of fact, the problem of large-scale context is a general coding problem rather than a Lean-specific one, so it is perhaps something that will improve independently of this field.
In the meantime, linters can help by enforcing the quality standards that are deterministic enough to be written down. 
Simple examples are the linters that check that every line is at most one hundred characters long, the standard for Mathlib code, or whether a declaration has unused arguments~\cite{mathlibStyleLinters,batteriesLinters}.
Such automated checks are cheap, and apply to AI-generated code exactly as they do to human-generated code.

Another common failure mode that would benefit from automatic detection is statements whose names sound like physical identities, but whose proof is just about unfolding a definition that already includes that property by construction. 
Taken by themselves, these statements are locally true and well-formed, but they do not establish that the intended physical identity has been derived.

\paragraph{Efficiency}

Formalizing results from mathematics and physics has historically been a notoriously difficult and slow process. 
Within Lean, a step that takes a line on paper can take a day to write, and historically, humans had to bear this cost. 
The efficiency with which results can be formalized determines how much of physics will ever be formalized and how many resources and people will be needed to do so. 

This is where AI speed has the most obvious effect. 
AI dramatically increases the amount of code that we can write, and it becomes a lot easier and a lot quicker for people to formalize results that they are interested in.

This comes with a downside, however: when a proof arrives quickly, there is very little pressure for anyone to read it. 
The fact that the code compiles creates overconfidence, and the user tends to forget to ask whether the code is readable, whether definitions and theorems have been stated in the right way and mean what they claim to mean, or whether they are equivalent to anything else in the project that could have been reused instead.

This is sometimes acceptable for one-off projects to quickly make progress. However, when such results are added, or proposed for addition, to larger libraries, the time saved by the author is often paid for by the reviewers and maintainers of those projects. 

\paragraph{Community}

The cost that efficiency shifts away from the author does not disappear. 
It falls on the community that reviews and maintains the result, and it is worth spelling out what that community actually does.

Indeed, a large part of formalization is service work. 
This includes taking a result that is already known to be true and making the best possible formalization of it, placing it in the right place within a project with the right documentation, and reviewing other people's contributions to make sure that everything is as good as possible.
It also includes onboarding new members and bumping the versions of upstream dependencies. This work is very similar to the service work that happens throughout open-source software development communities.

Previously, the people doing service work were credited indirectly, because they were also formalizing results and putting them into Lean, which is what the wider community recognized. However, this is exactly the part that AI now makes cheap. 
The people doing service work now no longer get the credit that they deserve, because service work absorbs most of their time and formalizing new results is cheap. 
This is a big problem in the community, because it undervalues the work that is most needed. 
For this reason, we think that incentives need to change, so that people get the credit they deserve for service work and the formalization ecosystem can thrive. 

In addition, there is a second effect, on the training of new contributors.
If AI does the work, the human skill naturally diminishes, since nobody needs the skill to get the result.
In the long run, we could end up in a situation where formalization can only be done with AI assistance because we never trained anybody to do it by hand.
What is lost then is not the ability to produce Lean code, but the understanding that comes from struggling with it. 
This understanding is vital for spotting nuances and challenges that matter for writing good code, and losing the judgment that goes into these decisions leads to worse code. 

\paragraph{Sustainability}

We believe that science is more effective when new results build on previous ones without having to rewrite everything from scratch, and projects like Mathlib and Physlib support this thesis.

Sustainability asks whether, for these sorts of projects, the community's capacity can keep up with the volume of work that is being produced. 
Historically, this has always been a problem. 
Mathlib has always had quite a large backlog of pull requests, even from human authors.
Ideally, to prevent the flooding of unchecked results that the community is expected to review, a contributor of a formalization should understand their contribution well enough to give an expert-level technical talk on the result~\cite{leiden2026declaration, tao2026mathematicsageai}, regardless of how they produced it.
Managing this backlog and making Mathlib sustainable eventually led to the creation of the Mathlib Initiative, which now hires people to help review pull requests and make the project more sustainable. 
Such a thing does not yet exist for Physlib. 

If used the right way, AI can actually help sustainability. 
It can be used to create tools that fit within the ecosystem, such as helping with the maintenance of projects, finding duplicates within existing code, finding generalizations, and finding unneeded arguments.
These are generally unglamorous things compared to the sort of end-to-end formalizations that lots of AI researchers usually target. 
The risk is that, because of the efficiency that AI gives in writing Lean code, there is a pull towards these more one-off sorts of demonstrations, taking resources away from the foundational projects that need them.

\paragraph{Applicability}
Applicability differs slightly from the dimensions discussed above because it concerns not the formalization itself, but whether a formalization is worth having in the first place. 

Indeed, there are some areas of academia where formalization is probably not the way forward. 
History is an example: it is governed by brute facts whose sequence and implications cannot be logically predicted, since every effect is the consequence of many causes that are hard to model.
In such a field, the value of formalization could actually be negative because it gives the impression of trust without actually providing it. 
The fact that AI could appear to formalize results in such fields is actually harmful, because when people turn to it in these areas, it may give a false certificate of value. 

Some fields are likely to benefit from formal methods first. 
Examples include mathematical or theoretical physics with heavy or contested derivations (e.g., quantum field theory, perturbation theory in general relativity, or quantum information), model search over large parameter spaces that should be physically grounded, and areas connected to computational methods, where theoretical algorithms are used to approximate physical quantities and then turned into code.

On the other hand, fields dominated by phenomenological models obtained by fits and approximations may appear less suitable for formalization. 
However, even in these cases, progress can be made. 
For example, it may be possible to describe and control approximations by making explicit the error-bound hypotheses, such as the asymptotic order and the regime inequalities. 
Physics lacks the habit of writing these details down, and formalization would make it mandatory. 
This habit could, for instance, prevent completely unphysical approximations that nevertheless fit the empirical data.

Historically, there was always the question of whether it was worth formalizing a given area. 
For example, it may take a lot of time and energy to formalize a specific problem in classical mechanics that nobody may ever use again.
As AI reduces the cost of making such formalizations, it is now worth formalizing many more problems than before. 
We see this as a great advantage.

\subsection{Near-term opportunities}
\label{subsec:AIForm:near-term}
In the short term, there are a few clear directions in which AI applications can broadly accelerate physics formalization.
These support the goals outlined in the Introduction: verifying physics results, making calculations reusable, and building the shared foundations needed for results across specified classes of models.

\paragraph{Planning}

A blueprint~\cite{leanblueprint} in Lean is a structured document that links mathematical text to formal declarations or results that will eventually be formalized. These have historically been used to plan large-scale formalizations, for example in the Fermat's Last Theorem project led by Kevin Buzzard.

Recently, new approaches to planning formalization have emerged, specifically designed with AI in mind~\cite{chen2026prove2meopencollaborativeplatform}. However, the general concept remains the same. Having a plan for what to formalize and how that relates to what has already been formalized makes formalization much easier for both AI and humans.

The project Tau Ceti has roadmaps for this, specifically designed for AI, and Physlib has a similar concept called API maps, designed for humans and AI within the context of PhyslibAlpha. 

The problem is that creating such documents requires an enormous amount of work. One has to track the dependencies of all the results within an area, work out how to convert text in the literature into something better suited to formalization, and make sure that the results in the blueprint represent not only what is in the code but also what is in the literature.

A near-term opportunity is to build AI tools to help with this. 
Planning is more important in physics than in mathematics, because in physics the hardest part of formalization is often working out exactly what to formalize in the first place. 
This is because physics results are not written in terms of definitions and theorems, as in pure mathematics, so extracting them often requires much more work. 

\paragraph{Reviewing contributions}
To compensate for the large increase in contributions enabled by AI, and to stimulate the growth of the physics formalization community by allowing new adopters to contribute and get feedback, it is necessary to find more efficient ways to review the contributions.
In particular, AI assistance in a preliminary assessment of a contribution to an open-source formal library, with agents specifically designed for physics formalization, would provide the first layer of review before an expert reviewer needs to intervene.
For example, these agents would compare proposed definitions to existing reference books, check which relevant properties have been verified, and detect possible ways in which a theorem may use unsound hypotheses or assumptions.

\paragraph{Shared community standards}
Much of the efficiency gain from AI currently depends on each user working out for themselves how to use general-purpose harnesses for formalization. 
Shared standards, skills, and tooling would remove this duplicated effort, so that the average user does not have to start from scratch, and would at the same time make the output more uniform and easier to review.

A harder but, we believe, unavoidable piece of this is writing down what we mean by quality in the first place. 
Linters can enforce the standards that are deterministic enough to state, but the deeper judgments, such as what makes a definition well designed, a file well organized, or a result stated at the right level of generality, currently live only in the heads of experienced contributors. 
Putting them into words is incredibly difficult, and an AI cannot be asked to meet a standard which has never been stated, and neither, in truth, can a new human contributor.
In physics, these standards differ from those used in mathematics formalization projects.

\paragraph{Reliable tooling}
AI agents can be effective when they are given reliable tools that implement specific low-level operations. Specialized software has the advantage of being more reliable and interpretable, and prevents the agent from rebuilding commonly used functions from scratch every time they are needed.
For example, solid definition and theorem search tools~\cite{gao2026leansearchv2globalpremise, loogle, Asher_LeanExplore_2025, lu2026leanfindersemanticsearch} are crucial to allow the agent to reuse the existing results and set up the proving strategy according to what is already available in the imported libraries. 
It is important to emphasize that theorem search is harder than simple keyword search because similar concepts can be expressed with different syntax, and also different fields may have similar and reusable mathematics. 
In addition, advanced access to the compiler can give high-quality feedback to the agent~\cite{leanLspMcp}. Being able to quickly get the compiler output or state and open goals at a given position of the code provides very relevant context to the agent.
Furthermore, tactics specialized for physics formalization that are recurrently required in several projects could become self-standing projects. 
Examples include index contraction and normalization for tensors~\cite{toobysmith2024indexnotation}, dimensionality checking~\cite{bobbin2025formalizingdimensionalanalysisusing}, symbolic asymptotic expansion and more advanced `simp' tactics.

\paragraph{Training and supporting material}
As with learning to code, there is the question of whether it is still worth learning to work with formal methods, and to what level, in a world where artificial intelligence can do most of the heavy lifting. 
Currently, it seems that at least a high-level understanding is important to be able to steer the AI agents in a formalization effort and to be able to trust the results. 
While AI can speed up the development, the more a researcher knows about formal methods, the easier it is to detect errors or loopholes, and eventually full trust is only possible if one can understand the full proof.

For this reason, to build a community of physicists who are capable of doing formal work, it will be necessary to develop appropriate supporting material, similar to that available for mathematics~\cite{AvigadMassot2020, lean4game}. 
Subsequently, for example, students can start small formalization projects about existing papers, starting from more accessible topics such as linearized general relativity perturbations and their stability, or gauge group representations~\cite{krippendorf2026physicscodescanstheorems}.
It can also be useful to try to rebuild results that are already formalized, for example those in existing open-problem benchmarks for physics formalization~\cite{yin2026axqmtextbookscalebenchmarkformal}.
In addition, exploring existing projects~\cite{PhysLib} and inspecting their definitions, theorems and proofs can also be very instructive.

\subsection{Long-term vision for physics}
As scientific practice evolves and integrates new technologies, the entire scientific workflow is being redesigned.
The systematic adoption of formal methods in physics can make theories much more consistent and precise, and even uncover new loopholes, especially in mathematical and theoretical physics.
Results can be shared and checked more easily and objectively, and each new result can be integrated into a common formal physics knowledge library.

Artificial intelligence offers paths to address some of the main bottlenecks that scientists, and physicists in particular, currently face in this process.
Even if their solution is not yet fully clear, we can outline some longer-term directions.

\paragraph{Verification and review of papers} 
One of the clearest long-term visions, which is certainly in a lot of people's minds, is that, in the future, papers in mathematically oriented areas will be accompanied by formal code verifying the results in that paper. 
This will not validate what has been done in the paper, but it will check the mathematical consistency of its results, as long as the Lean code accurately represents the paper itself, something which must also be checked. 
We believe that this will become a common first check on submitted work in these areas, and AI can really help to lower the entry barrier for researchers. 

\paragraph{Organization of scientific knowledge}
The above point assumed that papers will remain the academic unit of dissemination. We think it is quite possible that this will no longer be the case. 
It is likely that, given the scale at which AI can produce results, we will find new ways to collect, store, and process scientific information that avoid duplication and are better suited to the new world that we live in. 
Many candidates have been proposed, for example knowledge graphs~\cite{chaudhri2025community}, and we think the formal libraries that exist today, such as Mathlib and Physlib, will form some sort of foundation of that framework. 
This is only possible now that artificial intelligence can facilitate the process and allow the large-scale adoption and cataloging of such results. 

One benefit of this new organization would be easier retrieval of information, making it easier to find results and to see what other results have been obtained in the surrounding area. 
The cost of formalizing a new result is expected to fall as the library grows: most of a physics derivation is reuse of standard lemmas, and once these exist the remaining sorries are exactly the kind of local, checkable task AI fills reliably.

Scaling up reusable content also supports the longer-term goal of restricting physical model spaces discussed in the Introduction. Shared definitions and intermediate results could be combined into classification results, no-go or no-lose theorems, and proofs of necessary consequences under explicit assumptions. Such arguments must cover the specified class of models, beyond constructing individual examples. AI can help formalize their ingredients, while the choice of model class and its physical interpretation require scientific judgment.

\paragraph{Education and accessibility of science}
On a different note, AI can help students and researchers learn to use formal methods and start applying them within their domain of expertise.
In addition, teaching theoretical physics with the support of a formal library can help remove ambiguities and support textbooks with precise descriptions of the objects being referred to.

\paragraph{Verifying derivations and theory--experiment interface}
Finally, formal methods could also be adopted to verify the model that connects an experimental procedure to a theoretical observable, making sure that what is being measured corresponds to the right parameter of the physical model, and that all assumptions, approximations and hypotheses are appropriately tracked.
Work in this direction has already begun~\cite{ugwuanyi2026leanbetformallyverifiedsurfacearea}.

\section{Formalization to support AI applications in physics}
\label{sec:FormforAI}
The previous section treated formalization as the goal and AI as the supporting tool. 
Here we focus on the opposite role. Formal methods can help verify AI-generated derivations, establish guarantees on algorithms and numerical results, and provide a formal library as the environment in which an AI can reason about physics.

AI has been used in a variety of ways in physics, including inference, solving variational problems, and solving inverse problems~\cite{carleo2019machinelearning,quantum-review}. 
Examples include learning probability distributions for parameter inference, representing trial wavefunctions for quantum many-body problems, and reconstructing physical parameters from observations or designing materials with specified properties. Further applications include approximating solutions to differential equations and using surrogate models to accelerate simulations.

More recently, large language models and agentic systems have been used as assistants in theoretical developments~\cite{bubeck2025earlyscienceaccelerationexperiments, pan_quantum_2025}, in combination with numerical simulation tools and literature search~\cite{marquardt-naegele-agentic-expl, wang2025discoveryunstablesingularities}.
They can help brainstorm solutions, write code for analysis, and propose formulas.
These approaches are also becoming very popular in physics~\cite{Parisi_2026, schwartz2026resummationcparametersudakovshoulder, Menzo:2026xat}.
An AI system can propose conjectures~\cite{Guevara:2026qzd}, derivations or algorithms~\cite{feng2026aletheiatacklesfirstproofautonomously,brenner2026solvingopenproblemtheoretical,romera-paredes_mathematical_2024}, or even symbolic manipulations through computer algebra systems, but it needs a signal indicating which proposals are sound~\cite{novikov2025alphaevolvecodingagentscientific}.
For natural-language physics, that signal is usually a human reviewer, and it can interact with the agents through human-in-the-loop systems.

Across these applications, physicists seek to understand what the models have learned and how their predictions or proposed solutions relate to physical concepts. This includes interpreting learned representations and identifying the mathematical structures or algorithms associated with them. The broader use of machine learning to obtain rigorous results in physics and mathematics is reviewed in~\cite{Gukov:2024buj}. Formalization can support this understanding by making proposed mathematical descriptions, their assumptions, and their consequences explicit and checkable. It can also establish properties of the associated algorithms and guarantees on their numerical results. These provide a basis for deciding when AI outputs can be trusted and reused in further scientific reasoning.

\subsection{Opportunities for using formal methods}

Formal methods can support both numerical computation and scientific reasoning, from checking individual results to providing feedback during discovery and organizing reusable knowledge.

\paragraph{Verification and trust}

\leavevmode
Benchmarks of AI reasoning in physics~\cite{chung2025theoreticalphysicsbenchmarktpbench,zhang2025physreasoncomprehensivebenchmarkphysicsbased,qiu2025phybenchholisticevaluationphysical} highlight continuing difficulties with open-ended research problems~\cite{zhu2026probingcriticalpointcritpt,wiemann2026discoverphysicsbenchmarkingllmsoutofthebox}. Evaluating the reasoning itself also remains an open problem~\cite{lightman2023letsverifystepstep,zheng-etal-2025-processbench,ma2026reliablefinegrainedevaluationnatural}.
While evaluating whether the final answer is correct is an option, more complex problems also require understanding how it is obtained, and process-aware approaches are preferred~\cite{gao2025testtimescalingtechniquestheoretical}.
Looking only at the plausibility of the final result is insufficient for trust, interpretability and reliable automated improvement~\cite{cornelio2025needverificationaidrivenscientific,shalyt2026aifolloweinsteinsfootsteps,grosso2026readyaidrivendiscoveryai}.

First, formal verification can be employed after a full derivation has been developed informally to provide a reliable consistency check. 
Once a claim is expressed formally, a proof can be checked against its definitions and assumptions. A failed proof attempt may reveal a gap or a need for clarification, but does not by itself show that the claim is false.
This increases trust in AI-assisted reasoning. The checking can be automated, although finding a proof and validating the formal statement may still require substantial work.
If AI keeps contributing to scientific results more and more autonomously, formal methods can force AI systems to be precise and rigorous about each step, reducing the possibility of hallucinated results.
By allowing scientific reasoning to be expressed as code, formal methods will help AI models verify their proposals, detect and address loopholes and contradictions, and even explore the literature consistently.

\paragraph{Supporting computational methods}
Formal methods are also becoming very popular for general code verification.
As more infrastructure is developed~\cite{barrett2026cslibleancomputerscience, klaus2026rusttoleanverificationpipelineai}, new applications to computational methods will become possible, for example proving the equivalence of two different computational approaches, or showing when an analytical model approximates a computational method well.
Code verification can help prove that the implementation matches the theoretical algorithm or calculation~\cite{ugwuanyi2026leanbetformallyverifiedsurfacearea}; formal proofs can also establish the convergence or stability of a numerical scheme.
In addition, AI agents could aggressively optimize computational bounds, enabling safe AI-driven algorithm optimization.

Guarantees can also concern numerical outputs. For example, Monte Carlo approximation has been verified in Isabelle/HOL, including its integrability and convergence under stated assumptions~\cite{hirata2023probabilistic}. Such guarantees are probabilistic: they bound the chance of an error, rather than make each sampled estimate exact. For physics-informed neural networks, $\partial$-CROWN certifies bounds on differential-equation residuals over a continuous domain~\cite{eiras2024errorcertification}. A bound on the solution error additionally requires assumptions about the equation and its initial or boundary conditions.

\paragraph{Automated scientific discovery}
Beyond checking finished derivations, formal methods provide feedback during the search itself.
In AI for mathematics, proof assistants act as a deterministic verification layer, and this hard boundary allows agents to aggressively explore the space of theorems while their failures remain detectable.
In physics, the soundness of a proposal is usually judged manually, by human experts, and this feedback is scarce and does not scale with the number of new claims that AI can now generate~\cite{grosso2026readyaidrivendiscoveryai,Gu:2026zbf}.
Numerical certificates can also provide feedback, for example by establishing whether a proposed approximation meets a specified error tolerance.
Algorithm discovery systems such as AlphaEvolve~\cite{novikov2025alphaevolvecodingagentscientific} or AI-Descartes~\cite{cornelio2023aidescartescombiningdata}, in combination with AI-driven theorem proving systems, could receive feedback and use it to iterate on proposals for improving existing simulation algorithms and, in the future, also invent new ones.
Even a sparse formal verification signal can give useful feedback~\cite{zhang2026physproveradvancingautomatictheorem}.

For this purpose, a formal physics library is also crucial. It will provide the supporting mathematical content for physical reasoning, serving as the environment in which future AI systems reason about a theory, formulate conjectures, derive consequences and even propose experiments, reusing intermediate results across related cases.
Such systems may reason formally step by step and use the verification signal for training or for inference-time search, iteratively updating their proposals according to the compiler feedback~\cite{kamoi-etal-2026-efficient}.
Especially in those areas that are currently inaccessible to simulation algorithms or numerical methods, an analytical statement can help make progress. If the model proposes it in \LaTeX{} code, or derives it in Mathematica, it may still be hard to track down the assumptions, simplifications and regimes of validity. A formal proof makes the mathematical assumptions and steps explicit~\cite{krippendorf2026physicscodescanstheorems}.

Feedback matters most for the questions of classification and completeness raised in the Introduction. In particular, generative models can sample the space of models and produce candidates, but sampling alone cannot cover all possibilities or provide evidence that a certain property is unavoidable.
In contrast, no-go statements dramatically reduce the search space, and establishing them requires a proof that controls the entire class of models that are being considered, making explicit the assumptions and scope.
Formal methods and proof assistants can help to establish whether a property holds throughout that class.
As scientific discovery becomes more demanding~\cite{park_papers_2023,harder-to-find-ideas}, the upfront cost of formalization may be justified by its support for systematic exploration of specified theory spaces. Combined with appropriate search algorithms, formal methods can help develop or improve models and theories.
As many fields of physics evolve to integrate AI in their workflows, the success of formal methods in mathematics is seen as an encouraging precedent for what could be done in physics.

\paragraph{Interpretable foundation models}

The goal of many foundation models in machine learning is to create a world model, i.e., a compact representation of reality.
Approaches include learning dynamics~\cite{Schmidhuber-worldmodels,schrittwieser_mastering_2020}, predictive representations~\cite{assran2025vjepa2selfsupervisedvideo}, or generative models of environments~\cite{hu2023gaia1generativeworldmodel,bruce2024geniegenerativeinteractiveenvironments,hafner2024masteringdiversedomainsworld,nvidia2025cosmosworldfoundationmodel}.
Understanding what these models have learned involves relating their internal representations to physical concepts, such as space and time~\cite{li2024emergentworldrepresentationsexploring,gurnee2024languagemodelsrepresentspace}. Where a mathematical interpretation or an associated algorithm can be specified, formalization can help check its properties and its relation to the model's outputs. It does not by itself establish that a proposed interpretation describes the internal representation.

In some sense, at least for quantitative sciences, formal libraries could become complementary to large-scale foundation models based on the real world, acting as an interpretable and trustworthy layer in which AI models can reason about physics, and connect to black-box results.
As formalization becomes more and more popular in the field of artificial intelligence, common physics libraries can make theoretical results available for checked reasoning and reuse. As new results are proposed, verified and added to the library, they can be reused immediately, generalizing to new settings where they apply.
The difficulty can decrease as more results are integrated into a unified library that systems can use to search existing knowledge, identify similarities and generate ideas, and propose new results that either expand the existing ones or help explain new scientific evidence. In other words, such a library may acquire good generalization properties when extended to new domains, because structures and similar mathematical objects can be reused.

\paragraph{Encoding future scientific theories}
In the future, black-box models driven by neural networks may become more efficient than a physical theory at approximating simulations and giving predictions for a given phenomenon. 
However, mathematical models can offer simpler and more interpretable descriptions.
Since the goal of science is also to provide explanations that humans can understand, the construction of simpler mathematical models based on a few fundamental laws still has great value, not only for their potential to generalize, but also for the understanding, intuition, and new ideas they provide.
Expressing such models formally would allow computers to reason about their assumptions and consequences, and enable their results to be checked and reused. This provides a longer-term role for formalization in developing and comparing candidate physical theories.

\subsection{Limitations}
While the big challenges in mathematics are often expressed by statements whose correctness should be proven, the conjecture space in physics is broader and harder to explore, because it requires stronger intuition about the most appropriate definitions to model a physical system.
The guarantees of formal methods concern specified mathematical statements under explicit assumptions.
For numerical results, their scope also depends on the domain, approximation errors, and any probabilistic qualifications. Guarantees about an idealized algorithm apply to its implementation only when the relevant numerical errors and implementation details are accounted for.
Formal methods help to check consequences of a theory's assumptions, but its validity in the real world must still be assessed empirically through experiments.

Moreover, intuition, creativity and connection to experimental evidence are often the core mechanisms that enable scientific progress. 
A formal physical theory may emphasize loopholes and boundary cases, but also conceal the crucial physical meaning.
For this reason, there will still be space for interpretation and intuition about the formal results, which will still be discussed informally, so that humans, and general-purpose AI systems, can better reason about the mathematical results and their consequences in the real world.

\section{Conclusion}

Coming from different backgrounds in formalization and AI, and combining academic and industry experience, we are all very excited about the opportunities offered by formal methods in physics and their integration with artificial intelligence in the future. 
Formal methods will provide more precise and rigorous theories and physical models, tracking assumptions and preventing inconsistencies, while also enabling systematic and trustworthy reasoning through artificial intelligence, potentially dramatically accelerating scientific progress, especially in more theoretical and computational areas.

On the other hand, we recognize that several challenges must be overcome to reach that future.
At the end of 2026, we are in a much better place than at the beginning of the year, or in 2025. 
We are not yet at the point where formalization is easy and practical enough for anyone to produce high-quality formal code for their proofs without knowing anything about Lean or formalization. 
In the long run, we think we may reach the point where formal methods belong to the researcher's toolkit in the same way as coding, mathematics, or statistics. In this perspective, we have outlined the steps that we think are needed to get there. 

Lean is currently the most widely adopted formal language, and 
it was for this reason that we chose to focus on Lean in this work.
However, for a given community, some of the features which exist in Lean, such as allowing classical logic and therefore non-computable functions, and proof irrelevance, may not be optimal, or, likewise, some features which might be missing in Lean might be vital for some projects. 
If the community migrates to another interactive theorem prover, large bodies of AI-generated tactic scripts, which exploit idiosyncratic tactic behavior, may be hard to port and need to be thoroughly analyzed and rewritten. 
On the other hand, proofs that express the mathematical idea elegantly, as a human would, may be easier to transfer, since the general ideas are more evident.

As we previously mentioned, the systematic adoption of formal methods in physics will also require novel community and academic incentives and recognition. 
If service work is not credited, and artificial intelligence is increasingly adopted without human supervision, the field may develop into fragmented applications that are too brittle to be reliably reused or built upon. 
If misaligned incentives leave no knowledgeable experts in the community, and nobody is interested in investing time in learning and teaching the technical details, the field will stagnate.

In general, for many formal projects, service work is currently being professionalized. 
For example, the recently established Mathlib Initiative now supports mathematicians whose role is solely to review and provide feedback on contributions to the project, and the Lean Focused Research Organization employs core maintainers.
On the other hand, from the academic perspective, the problem of academic recognition is still open. 
As a matter of fact, formal methods allow scaling up collaboration on theoretical projects, and it is therefore reasonable to expect the adoption of attribution models similar to those used in large-scale experimental collaborations such as those at CERN~\cite{schukraft2006cernworkingexampleglobal}.
The Mathlib project, and more generally the formal mathematics community, partially address this for mathematics by attributing collective authorship~\cite{mathlib}, publicly naming maintainer and reviewer roles, and, more generally, listing every contributor to a community project as a coauthor~\cite{bolan2025equationaltheoriesprojectadvancing}.
However, the tendency to generalize concepts in Mathlib often requires deleting duplicated material to implement a more general version and, as a consequence, contributors' attributions may not persist in the library after each refactoring.

We believe that similar incentives should be established in the formal physics community.
At the moment, the community is smaller, has not yet been supported by nonprofit projects, and is still of marginal interest to many industrial efforts, which mainly focus on mathematics and code verification~\cite{achim2025aristotleimolevelautomatedtheorem,axiommath2026primegaps,hariharan2026progressformalizingspherepacking,requena2026axproverbase}.
Specifically, we believe that the community could adopt similar collective ``Physlib'' authorship approaches: making library releases citable by assigning them DOIs, and attributing authorship to everyone who contributed to the project, while also recognizing specific roles through the CRediT taxonomy~\cite{brand2015beyond}. 
Each contribution could be tracked and recognized through platforms 
like Palomar~\cite{palomar_registry_2026}, as in other formal projects.

In addition, at the academic level, we will need to better recognize formal library developers and maintainers, whose work should be considered of similar value to independent research outputs, both for hiring and for funding and grant assignment. 
In mathematics, a library itself can be the target of a grant~\cite{fltProject}.

Finally, AI-assisted results should be disclosed, and credit should be attached mainly to quality, comprehension, and high-level understanding rather than to the volume of results.

\section*{Data Availability}
No new data were generated or analyzed in this study.
\par\vspace*{0.3cm}

\section*{Acknowledgments}
We thank Xi Yin and Michael Douglas for helpful discussions in the earlier phase of this project.
L.S. thanks Dirk Englund for discussions. We thank Benjamin Breen and Matteo Cipollina for comments on the manuscript. 
This work was performed in part at the Aspen Center for Physics, which is supported by National Science Foundation grant PHY-2210452. S.K. has been partially
supported by STFC consolidated grants ST/T000694/1 and ST/X000664/1.

We disclose that LS works for Axiomatic AI which develops AI tools for formalization with industrial applications to physics, 
and that JTS is a maintainer and initiator of the project Physlib. 

\section*{AI declaration}
During the creation of this manuscript, the following AI tools were used to help with wording and grammar, and to write the scripts that create the figures: Claude Opus 5.5, GPT-6 Astra and GPT-5.6 Sol.
All thoughts and opinions in this article originated with the authors. Where AI was used to support the writing, the resulting text was carefully read and edited to ensure that it conveyed the original meaning that the authors intended. 
\bibliographystyle{apsrev4-2}
\bibliography{MyBib}

\end{document}